\documentclass[nofootinbib,prd,showpacs]{revtex4}
\usepackage{amsmath}
\usepackage{subfigure}
\usepackage{amsfonts}
\usepackage{hyperref}
\usepackage{amssymb}
\usepackage{graphicx}
\graphicspath{ {figs/} }

\usepackage[latin1]{inputenc}
\usepackage[english]{babel}
\usepackage{color,xcolor}

\definecolor{purple}{rgb}{1,0,1}
\definecolor{lime}{HTML}{A6CE39} 

\begin{document}
\title{
Embedding regular black holes and black bounces in a cloud of strings
}
\author{Manuel E. Rodrigues}
\email{esialg@gmail.com}
\affiliation{Faculdade de Ci\^{e}ncias Exatas e Tecnologia, 
Universidade Federal do Par\'{a}\\
Campus Universit\'{a}rio de Abaetetuba, 68440-000, Abaetetuba, Par\'{a}, 
Brazil}
\affiliation{Faculdade de F\'{\i}sica, Programa de P\'{o}s-Gradua\c{c}\~ao em 
F\'isica, Universidade Federal do 
 Par\'{a}, 66075-110, Bel\'{e}m, Par\'{a}, Brazil}
\author{Marcos V. de S. Silva}
\email{marco2s303@gmail.com}
\affiliation{Faculdade de F\'{\i}sica, Programa de P\'{o}s-Gradua\c{c}\~ao em 
F\'isica, Universidade Federal do 
 Par\'{a}, 66075-110, Bel\'{e}m, Par\'{a}, Brazil}
\date{today; \LaTeX-ed \today}
\begin{abstract}
A string is the one-dimensional generalization of a point particle. In this sense, the analog of a cloud of dust would be a cloud of strings. In this work, we consider a cloud of strings around a regular solution in general gelativity. We consider the Bardeen solution and the Simpson--Visser solution, analyzing the consequences of the cloud in the regularity and in the energy conditions. Actually, the presence of the cloud could make the energy density positive when compared to the Simpson--Visser case. We verify that the usual Bardeen solution becomes singular in the presence of the cloud of string while the Simpson--Visser solution is still regular, violating the energy conditions, as the usual solution. We also calculate some thermodynamic quantities to evaluate how a cloud of strings influences the thermodynamics of the solutions.
\bigskip

\end{abstract}
\maketitle
\def\HMS{{\scriptscriptstyle{HMS}}}
\section{Introduction}
\label{S:intro}
General relativity is a theory of gravitation where the gravitational interaction is a result of the curvature of the spacetime \cite{din,wal}. Since it was proposed, this theory has been tested and proved to be effective in describing phenomena that were already known before its creation, such as the procession of the perihelion of mercury \cite{Einstein:1915bz,Kraniotis:2003ig,Will:2018mcj}, as well as predicting new phenomena, such as the bending of light rays and gravitational waves \cite{din}. These predictions have been experimentally proven, with the bending of light being tested in 1919 \cite{Crispino:2019yew} and gravitational waves in 2015 \cite{LIGOScientific:2016aoc}. The detection of gravitational waves represents a major advance in the study of our universe, allowing us to obtain information from astrophysical objects that are located very distant from our solar system.

Another very relevant prediction of general relativity is the black hole \cite{Chandrasekhar:1985kt}. These astrophysical objects stand out due to their causal structure since from a certain region, known as the event horizon, no particle can escape from it \cite{wal}. Through the detection of gravitational waves, information was obtained from a vast number of black holes, some with small masses approaching $5$ solar masses and some with large masses of up to more than $100$ solar masses \cite{LIGOScientific:2018mvr,LIGOScientific:2020ibl,LIGOScientific:2021psn}.

A structure that is often present in black hole solutions is the singularity \cite{Stoica:2014tpa}. These structures are characterized by the fact that geodesics are interrupted by them \cite{Bronnikov:2012wsj}. In some cases, the presence of singularities can be verified through the curvature invariants \cite{Bronnikov:2012wsj}; these are known as curvature singularities. What characterizes a black hole is the presence of an event horizon \cite{wal}. In this way, it is possible for solutions without singularities to exist, which are the regular black holes \cite{Ansoldi:2008jw}. The first regular solution was proposed by Bardeen and interpreted as a solution of the Einstein equations in the presence of nonlinear electrodynamics by Ayon--Beato and Garcia \cite{Bardeen,Beato1}. There are a lot of regular solutions and studies on properties of these solutions in the literature \cite{Fan-Wang,Zaslavskii,Rodrigues:2015,Rodrigues:2017,Rodrigues:2018,Rodrigues:2019,Silva:2018,Junior:2020,Bronnikov:2017tnz,NED2,NED3,NED4,NED5,Bronnikov:2006fu,NED10,neves,toshmatov,ramon,berej}.

Another class of regular black holes was proposed by Simpson and Visser, solutions known as black bounces \cite{Simpson:2018tsi}. This solution distinguishes it from standard regular black holes through a modification in the black hole's area, allowing the presence of a nonzero radius throat at $r=0$. Initially, Simpson and Visser did not propose the content of material that could describe this solution. Recently, some works have appeared to show that solutions of black bounces can be obtained through the Einstein equations when there is a scalar field and nonlinear electrodynamics \cite{Bronnikov:2021uta,Canate:2022gpy}. In addition to the Simpson and Visser solution, there is a vast number of black bounces, where some of them have modifications in the area, and there are works focused on analyzing properties of these solutions \cite{Simpson:2019cer,Lobo:2020ffi,Simpson:2021vxo,Lobo:2020kxn,Tsukamoto:2021caq,Olmo:2021piq,Guerrero:2022qkh,Yang:2021cvh,Lima:2020auu,Lima:2021las,Huang:2019arj,Rodrigues:2022mdm}.

In 1978, Letelier proposed a new type of black hole solution. In this case, we have a standard Schwarzschild black hole surrounded by a cloud of strings \cite{Letelier:1979ej}. The cloud as a whole is a closed system, such that the stress-energy tensor is conserved \cite{Letelier:1979ej}. Several works analyze the consequences of the presence of the string cloud in different situations, such as accretion, thermodynamics, and quasinormal modes, and there are several proposed solutions considering the presence of the cloud \cite{Glass:1997qz,Ganguly:2014cqa,deMToledo:2018tjq,Toledo:2018hav,MoraisGraca:2016hmv,DiaseCosta:2018xyj,Chabab:2020ejk,Cai:2019nlo,Ghosh:2014pga,Ghosh:2014dqa}. There is a solution where the authors consider a regular black hole surrounded by the cloud; however, due to the presence of the cloud, the solution is not regular anymore \cite{Sood:2022fio}. It would be interesting if there were regular solutions whose presence of the string cloud does not insert a singularity. This is one of the purposes of this work.

The structure of this work is organized as follows. In section \ref{S:general} we presented the field equations to a cloud of strings, considering a general spacetime. In section \ref{S:energyconditions} we presented the energy condition to a general spacetime, which will be needed in the following sections. In section \ref{thermo} we present the thermodynamics quantities that are necessary to study the thermodynamic equilibrium of a solution. The Bardeen solution surrounded by the cloud of strings is present in section \ref{BardeenString}, having its regularity and energy conditions analyzed. In section \ref{SimpsonCloud} we propose a stress-energy tensor that allows us to obtain the Simpson--Visser solution in a cloud of string. We also analyzed the regularity of this solution and the energy conditions. Our conclusions are present in section \ref{S:conclusion}.

 We adopt the metric signature $(+,-,-,-)$.
 Given the Levi-Civita connection, $\Gamma^{\alpha}{}_{\mu\nu}=\frac{1}{2} g^{\alpha\beta}\left(\partial_{\mu}g_{\nu\beta}+\partial_{\nu}g_{\mu\beta}-\partial_{\beta}g_{\mu\nu}\right)$, the Riemann tensor is defined as $R^{\alpha}{}_{\beta\mu\nu}=\partial_{\mu}\Gamma^{\alpha}{}_{\beta\nu}-\partial_{\nu}\Gamma^{\alpha}{}_{\beta\mu}+\Gamma^{\sigma}{}_{\beta\nu}\Gamma^{\alpha}{}_{\sigma\mu}-\Gamma^{\sigma}{}_{\beta\mu}\Gamma^{\alpha}{}_{\sigma\nu}$. We shall work in geometrodynamics units where $G=\hbar=c=1$. 
\bigskip

\clearpage
\section{Spacetimes With Cloud of strings}
\label{S:general}
We are interested in obtaining solutions that describe spacetimes in the presence of a cloud of strings. In this way, we will analyze these solutions in the context of general relativity. The action that describes general relativity minimally coupled with matter and the cloud of strings is given by 
\begin{equation}
    S=\int d^4x \sqrt{-g}R+S_M+S_{CS},\label{Actiongeral}
\end{equation}
where $S_M$ is the action that describes the matter sector, which will be specified later, and $S_{CS}$ is the Nambu--Goto action used to describe stringlike objects, given by \cite{Letelier:1979ej}
\begin{equation}
    S_{CS}=\int \sqrt{-\gamma}\mathcal{M}d\lambda^0d\lambda^1.\label{ActionCS1}
\end{equation}
Here, $\mathcal{M}$ is a dimensionless constant that characterizes the string and $\gamma$ is the determinant of $\gamma_{AB}$, which is an induced metric on a submanifold given by
\begin{equation}
    \gamma_{AB}=g_{\mu\nu}\frac{\partial x^\mu}{\partial \lambda^A}\frac{\partial x^\nu}{\partial \lambda^B}.
\end{equation}
Just as a particle is associated with a world line, a string is associated with a world sheet, which is described by $x^\mu(\lambda^A)$, where $\lambda^0$ and $\lambda^1$ are timelike and spacelike parameters.

It is possible to write the Nambu--Goto action as
\begin{equation}
    S_{CS}=\int \mathcal{M}\left(-\frac{1}{2}\Sigma^{\mu\nu}\Sigma_{\mu\nu}\right)^{1/2}d\lambda^0d\lambda^1,
\end{equation}
where $\Sigma^{\mu\nu}$ is a bivector written as
\begin{equation}
    \Sigma^{\mu\nu}=\epsilon^{AB}\frac{\partial x^\mu}{\partial \lambda^A}\frac{\partial x^\nu}{\partial \lambda^B},
\end{equation}
with $\epsilon^{AB}$ being the Levi--Civita symbol, $\epsilon^{01}=-\epsilon^{10}=1$.

Varying the action \eqref{Actiongeral} with respect to the metric we find
\begin{equation}
    R_{\mu\nu}-\frac{1}{2}Rg_{\mu\nu}=\kappa^2T_{\mu\nu}=\kappa^2T^{M}_{\mu\nu}+\kappa^2T^{CS}_{\mu\nu}.\label{Einstein}
\end{equation}
Here, $T^{M}_{\mu\nu}$ and $T^{CS}_{\mu\nu}$ are the stress-energy tensor of the matter sector and the stress-energy tensor of the cloud of string. We will only define the form of $T^{M}_{\mu\nu}$ when considering specific cases. The form of $T^{CS}_{\mu\nu}$ is \cite{Letelier:1979ej}
\begin{equation}
    T^{CS}_{\mu\nu}=\frac{\rho\Sigma_{\mu}^{\ \alpha}\Sigma_{\alpha\nu}}{8\pi\sqrt{-\gamma}},
\end{equation}
where $\rho$ is the proper density of the cloud. The stress-energy tensor must obey the conservation law
\begin{equation}
    \nabla_\mu {T^{CS}}^{\mu\nu}=\nabla_\mu\left(\frac{\rho\Sigma^{\mu \alpha}{\Sigma_{\alpha}}^\nu}{8\pi\sqrt{-\gamma}}\right)=\nabla_\mu\left(\rho\Sigma^{\mu \alpha}\right)\frac{{\Sigma_{\alpha}}^\nu}{8\pi\sqrt{-\gamma}}+\rho\Sigma^{\mu \alpha}\nabla_\mu\left(\frac{{\Sigma_{\alpha}}^\nu}{8\pi\sqrt{-\gamma}}\right)=0.
\end{equation}
Multiplying the equation above by $\Sigma_{\nu\beta}/(-\gamma)^{1/2}$ and using the identity
\begin{equation}
    \Sigma^{\alpha\beta}\nabla_\alpha\Sigma_{\beta\nu}=\frac{3}{2}\Sigma^{\alpha\beta}\partial _{[\alpha}\Sigma_{\beta\nu]}-\frac{1}{4}\nabla_\nu\left(\Sigma_{\nu\beta}\Sigma_{\alpha\beta}\right),
\end{equation}
it is possible to obtain the following equations \cite{Letelier:1979ej}:
\begin{equation}
    \nabla_\mu \left(\rho \Sigma^{\mu\nu}\right)= \partial_\mu \left(\sqrt{-g}\rho \Sigma^{\mu\nu}\right)=0,\label{conserv1}
\end{equation}
\begin{equation}
    \Sigma^{\mu\beta}\nabla_\mu\left[{\Sigma_{\beta}}^{\nu}/(-\gamma)^{1/2}\right]=0.\label{conserve2}
\end{equation}

A general line element that describes a spherically symmetric and static spacetime is written as
\begin{equation}
    ds^2=A(r)dt^2-B(r)dr^2-C(r)\left(d\theta^2+\sin^2\theta d\phi^2\right).
\end{equation}
However, to regular black holes, in the context of the general relativity, we usually have $A(r)=B(r)^{-1}$ and $C(r)=r^2$. When considering black bounces we still have $A(r)=B(r)^{-1}$ but $C(r)\neq r^2$. Therefore, it is useful for us to use the ``Buchdahl coordinates'', where the line element is
\begin{equation}
    ds^2=f(r)dt^2-f(r)^{-1}dr^2-\mathcal{R}^2\left(d\theta^2+\sin^2\theta d\phi^2\right),\label{ele}
\end{equation}
where $f(r)$ and $\mathcal{R}$ are general functions of the radial coordinate.

Using the line element \eqref{ele}, we can integrate \eqref{conserv1} and \eqref{conserve2}, and it results in
\begin{equation}
    \Sigma^{01}=\sqrt{-\gamma}=\frac{a}{\rho \mathcal{R}^2},\label{sigma01}
\end{equation}
where $a$ is an integration constant related with the string. The nonzero components of the stress-energy tensor are
\begin{equation}
    {{T^{CS}}^{0}}_0={{T^{CS}}^{1}}_1=\frac{a}{8\pi \mathcal{R}^2}.\label{SETstring}
\end{equation}
In order to obtain a positive energy density, $a>0$.

The nonzero components of the Riemann tensor are given by
\begin{equation}
    {R^{01}}_{01}=\frac{1}{2}f', \quad {R^{02}}_{02}={R^{03}}_{03}=\frac{f'\mathcal{R}'}{2\mathcal{R}},\quad {R^{12}}_{12}={R^{13}}_{13}=\frac{f'\mathcal{R}'+2f\mathcal{R}''}{2\mathcal{R}},\quad {R^{23}}_{23}=\frac{f\mathcal{R}'^2-1}{\mathcal{R}}.
\end{equation}
The Kretschmann scalar is written as a sum of squares of the Riemann tensor components
\begin{eqnarray}
    K&=& 4\left({R^{01}}_{01}\right)^2+4\left({R^{02}}_{02}\right)^2+4\left({R^{03}}_{03}\right)^2+4\left({R^{12}}_{12}\right)^2+4\left({R^{13}}_{13}\right)^2+4\left({R^{23}}_{23}\right)^2\nonumber\\
    &=&4\left({R^{01}}_{01}\right)^2+8\left({R^{02}}_{02}\right)^2+8\left({R^{12}}_{12}\right)^2+4\left({R^{23}}_{23}\right)^2.\label{KreRiemann}
\end{eqnarray}
If there is divergence in any of the components of the Riemann tensor, this divergence must also appear in the Kretschmann scalar \cite{Lobo:2020ffi,Bronnikov:2012wsj}.
To the line element \eqref{ele}, \eqref{KreRiemann} is given by
\begin{equation}
K=\frac{
(\mathcal{R}^2 f'')^2+2(\mathcal{R} f' \mathcal{R}')^2 +2\mathcal{R}^2(f' \mathcal{R}'+2f \mathcal{R}'')^2
+4(1-f\mathcal{R}'^2)^2}{\mathcal{R}^4}\,.
\label{Kret}
\end{equation}
If Eq. \eqref{Kret} is singular in any point, the spacetime presents curvature singularities \cite{Lobo:2020ffi,Bronnikov:2012wsj}.

Now we need to obtain the form of $f(r)$ and $\mathcal{R}(r)$ but it is still necessary to define the form of $S_M$.
\section{Energy Conditions}\label{S:energyconditions}
In general, any models of $f(r)$ and $\mathcal{R}$ can be proposed. However, these models are not always physically appropriate. So, to know if the solutions make physical sense, we need to check the energy conditions.

To obtain the energy conditions, we need to identify the components of the stress-energy tensor. Where $t$ is the timelike coordinate, $f>0$, we have
\begin{equation}
    {T^{\mu}}_{\nu}={\rm diag}\left[\rho,-p_r,-p_t,-p_t\right],
\end{equation}
where $\rho$ is the energy density, $p_r$ is the radial pressure, and $p_t$ is the tangential pressure. Considering the Einstein equations and Eq. \eqref{ele}, we find
\begin{eqnarray}
&&\rho=-\frac{\mathcal{R} \left(f' \mathcal{R}'+2 f \mathcal{R}''\right)+f \mathcal{R}'^2-1}{\kappa ^2 \mathcal{R}^2}\label{density}\,,\\
&&p_r=\frac{\mathcal{R} f'\mathcal{R}'+f \mathcal{R}'^2-1}{\kappa ^2 \mathcal{R}^2}\,,\label{pr}\\
&&p_t=\frac{\mathcal{R} f''+2 f' \mathcal{R}'+2 f \mathcal{R}''}{2 \kappa ^2 \mathcal{R}}\label{pt}\,.
\end{eqnarray}

To regions where $f(r)<0$, $t$ is the spacelike coordinate, we must have
\begin{eqnarray}
T^{\mu}{}_{\nu}={\rm diag}\left[-p_r,\rho,-p_t,-p_t\right]\,.\label{EMT2}
\end{eqnarray}
The fluid quantities, in this region, are
\begin{eqnarray}
&&\rho=-\frac{\mathcal{R} f'\mathcal{R}'+f \mathcal{R}'^2-1}{\kappa ^2 \mathcal{R}^2}\label{density2}\,,\\
&&p_r=\frac{\mathcal{R} \left(f' \mathcal{R}'+2 f \mathcal{R}''\right)+f \mathcal{R}'^2-1}{\kappa ^2 \mathcal{R}^2}\,,\label{pr2}\\
&&p_t=\frac{\mathcal{R} f''+2 f' \mathcal{R}'+2 f \mathcal{R}''}{2 \kappa ^2 \mathcal{R}}\label{pt2}\,.
\end{eqnarray}

The energy conditions~\cite{book} are given by the inequalities
\begin{eqnarray}
&&NEC_{1,2}=WEC_{1,2}=SEC_{1,2} 
\Longleftrightarrow \rho+p_{r,t}\geq 0,\label{Econd1} \\
&&SEC_3 \Longleftrightarrow\rho+p_r+2p_t\geq 0,\label{Econd2}\\
&&DEC_{1,2} \Longleftrightarrow \rho-|p_{r,t}|\geq 0 \Longleftrightarrow 
(\rho+p_{r,t}\geq 0) \hbox{ and } (\rho-p_{r,t}\geq 0),\label{Econd3}\\
&&DEC_3=WEC_3 \Longleftrightarrow\rho\geq 0.\label{Econd4}
\end{eqnarray}
We see that $DEC_{1,2} \Longleftrightarrow ( (NEC_{1,2}) \hbox{ and } (\rho-p_{r,t}\geq 0))$, 
so we replace $DEC_{1,2}\Longrightarrow \rho-p_{r,t}\geq 0$.

Inserting the results given in  \eqref{density}--\eqref{pt}, where the coordinate $t$ is timelike, we have

\begin{eqnarray}
&&NEC_{1}=WEC_1=SEC_1 \Longleftrightarrow
-\frac{2 f \mathcal{R}''}{\kappa ^2 \mathcal{R}}\geq 0,\label{cond1+}\\
&&NEC_2=WEC_2=SEC_2 \Longleftrightarrow
\frac{\mathcal{R}^2 f''-2 f \left(\mathcal{R} \mathcal{R}''+(\mathcal{R}')^2\right)+2}{2 \kappa ^2 \mathcal{R}^2}\geq 0,\label{cond2+}\\
&&SEC_3 \Longleftrightarrow
\frac{\mathcal{R} f''+2 f' \mathcal{R}'}{\kappa ^2 \mathcal{R}}\geq0,\label{cond3+}\\
&&DEC_{1} \Longrightarrow
\frac{2\left(1 - f' \mathcal{R}\mathcal{R}' - f (\mathcal{R}')^2 -f \mathcal{R}\mathcal{R}'' \right)}{
	\kappa^2\mathcal{R}^2}\geq 0,\label{cond4+}\\
&&DEC_2 \Longrightarrow
-\frac{\mathcal{R}^2 f''+\mathcal{R} \left(4 f' \mathcal{R}'+6 f \mathcal{R}''\right)+2 f \mathcal{R}'^2-2}{2 \kappa ^2 \mathcal{R}^2}\geq0, \label{cond5+}\\
&&DEC_3=WEC_3 \Longleftrightarrow
-\frac{\mathcal{R} \left(f' \mathcal{R}'+2 f \mathcal{R}''\right)+f (\mathcal{R}')^2-1}{\kappa ^2 \mathcal{R}^2}\geq0. \label{cond6+}
\end{eqnarray}

To $f(r)<0$, $t$ is the spacelike coordinate, and the energy conditions are
\begin{eqnarray}
&&NEC_{1}=WEC_1=SEC_1 \Longleftrightarrow
+\frac{2 f \mathcal{R}''}{\kappa ^2 \mathcal{R}}\geq 0,\label{cond1-}\\
&&NEC_2=WEC_2=SEC_2 \Longleftrightarrow
\frac{ \mathcal{R}^2 f'' - 2 (\mathcal{R}')^2 f +2 \mathcal{R} \mathcal{R}'' f +2 }{2\kappa^2\mathcal{R}^2}\geq 0,
\label{cond2-}\\
&&SEC_3 \Longleftrightarrow
\frac{ \mathcal{R} f''+ 2 \mathcal{R}' f' + 4 \mathcal{R}'' f }{\kappa^2\mathcal{R}} \geq0,
\label{cond3-}\\
&&DEC_{1} \Longrightarrow
\frac{2\left(1 - f' \mathcal{R}\mathcal{R}' - f (\mathcal{R}')^2 -f \mathcal{R}\mathcal{R}'' \right)
	}{\kappa^2\mathcal{R}^2}\geq 0,
\label{cond4-}\\
&&DEC_2 \Longrightarrow
\frac{-\mathcal{R}^2 f'' - 2 \mathcal{R} \mathcal{R}'' f  -4 \mathcal{R}\mathcal{R}' f' -2 (\mathcal{R}')^2 f +2
	}{2\kappa^2\mathcal{R}^2} \geq 0,
\label{cond5-}\\
&&DEC_3=WEC_3 \Longleftrightarrow
-\frac{\mathcal{R} \mathcal{R}' f' + (\mathcal{R}')^2 f - 1 
	}{\kappa^2\mathcal{R}^2} \geq 0.
\label{cond6-}
\end{eqnarray}

\section{Thermodynamics}\label{thermo}
The thermodynamics of a black hole can give us information about the stability of a solution. To verify this stability, we calculate the Hawking temperature, which is given through the surface gravity as
\begin{equation}
T=\frac{k}{2\pi}=\left.\frac{f'(r)}{4\pi}\right|_{r=r_+},\label{Temp1}
\end{equation}
where $k$ is the surface gravity and $r_+$ is the radius of the event horizon.

In addition to temperature, a black hole also has an entropy associated with it. This entropy is given by the Bekenstein relation
\begin{equation}
 S=\frac{A}{4\pi}=\left.\pi \mathcal{R}^2\right|_{r=r_+}.   
\end{equation}

Once we have entropy and temperature, it is possible to calculate the heat capacity at constant charge through the relation
\begin{equation}
    C_q=\left.T\frac{\partial S}{\partial T}\right|_q.
\end{equation}
If the heat capacity is positive, the solution is thermodynamically stable \cite{Azreg-Ainou:2012zyx}.

\section{Bardeen solution with cloud of string}\label{BardeenString}
In General Relativity, the Bardeen solution arises when the gravitational theory is coupled with nonlinear electrodynamics. The action associated with the nonlinear electrodynamics is given by
\begin{equation}
    S_M=\int d^4x\sqrt{-g} \mathcal{L}(F),
\end{equation}
where $\mathcal{L}(F)$ is the Lagrangian density, which is a general function of the electromagnetic scalar $F=F^{\mu\nu}F_{\mu\nu}/4$. Here, $F_{\mu\nu}$ is the Maxwell-Faraday tensor. To a magnetically charged solution, the Lagrangian density to the Bardeen case is \cite{Beato1}
   \begin{equation}
\mathcal{L}(F)=\frac{3}{2sq^2}\left(\frac{\sqrt{2q^2F}}{1+\sqrt{2q^2F}}\right)^{5/2},\label{Lbardeen}
\end{equation}
where $s=\left|q\right|/2m$, with $q$ being the magnetic charge. The magnetic field and the electromagnetic scalar are given by
\begin{eqnarray}
F_{23}=q\sin\theta,\\
F=\frac{q^2}{2r^4}.\label{Fbardeen}
\end{eqnarray}

The stress-energy tensor to a nonlinear electrodynamics is
\begin{equation}
    T_{\mu\nu}=\frac{1}{4\pi}\left[g_{\mu\nu}\mathcal{L}-\mathcal{L}_F{F_{\mu}}^{\alpha}{F_{\nu\alpha}}\right],\label{STEelectro}
\end{equation}
with $\mathcal{L}_F=\partial \mathcal{L}/\partial F$.

Using \eqref{SETstring} and \eqref{STEelectro} with the Einstein equations, the equations of motion are
\begin{eqnarray}
&&\frac{a-\mathcal{R} \left(f' \mathcal{R}'+2 f \mathcal{R}''\right)-f \mathcal{R}'^2-2 \mathcal{L} \mathcal{R}^2+1}{\mathcal{R}^2}=0,\label{eqmov1}\\
&&\frac{a-\mathcal{R} f' \mathcal{R}'-f \mathcal{R}'^2-2 \mathcal{L} \mathcal{R}^2+1}{\mathcal{R}^2}=0,\label{eqmov2}\\
&&-\frac{f' \mathcal{R}'+f \mathcal{R}''}{\mathcal{R}}-\frac{f''}{2}-\frac{2 q^2 \mathcal{L}'}{\mathcal{R}^4 F'}-2 \mathcal{L}=0\label{eqmov3}.
\end{eqnarray}
Subtracting Eq. \eqref{eqmov1} from Eq. \eqref{eqmov2} we find
\begin{equation}
    -\frac{2 f \mathcal{R}''}{\mathcal{R}}=0.
\end{equation}
In general $\mathcal{R}(r)$ is not infinity and $f(r)$ is not zero, so that $\mathcal{R}''=0$, which means that
\begin{equation}
    \mathcal{R}=c_1 r+c_0,
\end{equation}
where $c_1$ and $c_0$ are integration constants. Imposing $c_0=0$ and $c_1=1$ we get that $\mathcal{R}=r$. 
Using Eq.\eqref{Lbardeen} and Eq. \eqref{Fbardeen}, Eq. \eqref{eqmov1} becomes
\begin{equation}
\frac{a}{r^2}-\frac{r f'(r)+f(r)-1}{r^2}-\frac{6 m \left(\frac{q^2}{q^2+r^2}\right)^{5/2}}{q^3}=0.    
\end{equation}
Solving this differential equation to $f(r)$, we obtain
\begin{equation}
    f(r)=1-a-\frac{2mr^2}{(r^2+q^2)^{3/2}},\label{fBardeen}
\end{equation}
where $a$ is limited by $0<a<1$. To $a=0$ the Bardeen solution is recovered \cite{Bardeen}. To $q=0$ the solution proposed by Letelier is recovered \cite{Letelier:1979ej}.

The Kretschmann scalar associated with the solution Eq. \eqref{fBardeen} is
\begin{equation}
    K=4 \left(\frac{a^2}{r^4}+\frac{4 a m}{r^2 \left(q^2+r^2\right)^{3/2}}+\frac{3 m^2 \left(-4 q^6 r^2+47 q^4 r^4-12 q^2 r^6+8 q^8+4 r^8\right)}{\left(q^2+r^2\right)^7}\right).
\end{equation}
Expanding the Kretschmann scalar to $r\rightarrow 0$ and $r\rightarrow \infty$, we find
\begin{eqnarray}
K &\approx& \frac{4 a^2}{r^4}+\frac{16 a m}{q^{3} r^2},\ \mbox{to} \ r\rightarrow 0,\\
K &\approx& \frac{4 a^2}{r^4}+\frac{16 a m}{r^5}+\frac{48 m^2}{r^6},\ \mbox{to} \ r\rightarrow \infty.
\end{eqnarray}
The singularity appears only at $r= 0$. If $a=0$, the therm that diverges disappears, so that there is curvature singularity only to  $a\neq 0$.

The energy conditions, where $t$ is the timelike coordinate, are given by

\begin{eqnarray}
&&NEC_{1} \Longleftrightarrow
 0, \qquad \qquad NEC_2 \Longleftrightarrow
\frac{a}{\kappa ^2 r^2}+\frac{15 m q^2 r^2}{\kappa ^2\left(q^2+r^2\right)^{7/2}}\geq 0,\label{Sol1cond1+}\\
&&WEC_3 \Longleftrightarrow
\frac{6 m q^2}{\kappa ^2\left(q^2+r^2\right)^{5/2}}+\frac{a}{\kappa ^2r^2}\geq0, \qquad \quad SEC_3 \Longleftrightarrow
\frac{6 m \left(3 q^2 r^2-2 q^4\right)}{\kappa ^2 \left(q^2+r^2\right)^{7/2}}\geq0,\label{Sol1cond2+}\\
&&DEC_{1} \Longrightarrow
 \frac{12 m q^2}{\kappa ^2\left(q^2+r^2\right)^{5/2}} +\frac{2a}{\kappa ^2r^2}\geq 0,\qquad \quad DEC_2 \Longrightarrow
-\frac{a}{\kappa^2r^2}+\frac{3 m q^2 \left(r^2-4 q^2\right)}{\kappa ^2\left(q^2+r^2\right)^{7/2}}\geq0. \label{Sol1cond3+}
\end{eqnarray}

Where $t$ is the spacelike coordinate, the energy conditions are
\begin{eqnarray}
&&NEC_{1} \Longleftrightarrow
 0,\qquad\qquad NEC_2\Longleftrightarrow
\frac{a}{\kappa ^2 r^2}+\frac{15 m q^2 r^2}{\kappa ^2 \left(q^2+r^2\right)^{7/2}}\geq 0,
\label{Sol1cond1-}\\
&&WEC_3 \Longleftrightarrow
\frac{a}{\kappa ^2r^2}+\frac{6 m q^2}{\kappa ^2\left(q^2+r^2\right)^{5/2}} \geq 0,\qquad \qquad SEC_3 \Longleftrightarrow
\frac{6 m \left(3 q^2 r^2-2 q^4\right)}{\kappa ^2 \left(q^2+r^2\right)^{7/2}} \geq0,
\label{Sol1cond2-}\\
&&DEC_{1} \Longrightarrow
\frac{2 a}{\kappa ^2r^2}+ \frac{12 m q^2}{\kappa ^2\left(q^2+r^2\right)^{5/2}}\geq 0, \qquad DEC_2 \Longrightarrow
\frac{a }{\kappa ^2 r^2}+\frac{3mq^2\left(4 q^2-r^2\right)}{\kappa ^2 \left(q^2+r^2\right)^{7/2}} \geq 0.
\label{Sol1cond3-}
\end{eqnarray}
We see that the null energy condition is satisfied, which means that the weak energy condition is also satisfied. The presence of the cloud of strings do not modify $SEC_3$, so the strong energy condition is violated inside the event horizon. The dominant energy condition is violated outside the event horizon.

\begin{figure}
    \centering
    \includegraphics[scale=0.9]{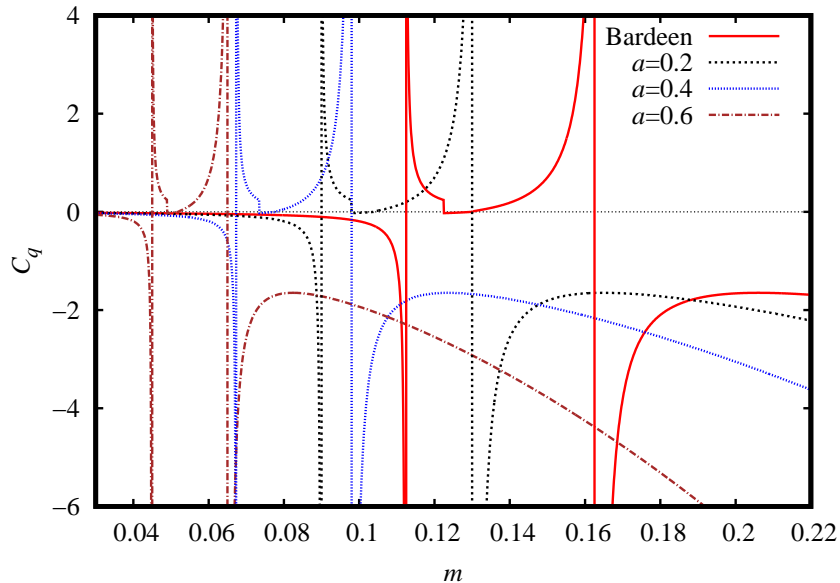}
    \caption{Heat capacity at constant charge to the solution Eq. \eqref{fBardeen} with $q=0.1$.}
    \label{fig:Cq1}
\end{figure}

From Eq. \eqref{Temp1}, the temperature is
\begin{equation}
T=\frac{m \left(r_+^3-2 q^2 r_+\right)}{2 \pi  \left(q^2+r_+^2\right)^{5/2}}. 
\end{equation}
However, from the condition $f(r_+)=0$, the black hole mass is given by
\begin{equation}
    m=\frac{(1-a) \left(q^2+r_+^2\right)^{3/2}}{2 r_+^2},\label{massBD}
\end{equation}
and the temperature becomes
\begin{equation}
    T=\frac{(1-a) \left(r_+^2-2 q^2\right)}{4 \pi  r_+ \left(q^2+r_+^2\right)}.\label{TempBD}
\end{equation}
The entropy to this case is
\begin{equation}
    S=\pi r_+^2.
\end{equation}
Once we have the temperature and entropy, the heat capacity at constant charge is
\begin{equation}
    C_q=\frac{2 \pi  r_+^2 \left(r_+^2-2 q^2\right) \left(q^2+r_+^2\right)}{7 q^2 r_+^2+2 q^4-r_+^4}.\label{HCBD}
\end{equation}
The constant $a$ does not appear in Eq.\eqref{HCBD}; however, we need to remember that $r_+=r_+(m,q,a)$. In Fig. \ref{fig:Cq1} see how the heat capacity behaves as the mass increases. There is an intermediate value of mass where the solution is stable, $C_q>0$. As the parameter $a$ increases, the range of mass where the solution is stable decreases.

\section{Simpson--Visser solution with cloud of strings}\label{SimpsonCloud}
The standard Simpson--Visser solution represents a regular black hole that has a minimum area that is not zero. Depending on the parameters of the metric, the solution can also represent a wormhole. To obtain the Simpson--Visser solution in the presence of a cloud of string, let us consider the Einstein equations \eqref{Einstein}, where ${T_{\mu\nu}}^M$ is

\begin{equation}
    T_{\mu\nu}^M=T_{\mu\nu}^{SV}+T_{\mu\nu}^{NMC},
\end{equation}
where $T_{\mu\nu}^{SV}$ is the stress-energy tensor associated with the Simpson--Visser solution, with the nonzero components given by \cite{Simpson:2018tsi}
\begin{eqnarray}
&&T^{SV}_{00}=-\frac{L^2 \left(\sqrt{L^2+r^2}-4 m\right) \left(\sqrt{L^2+r^2}-2 m\right)}{8\pi\left(L^2+r^2\right)^3}, \qquad T^{SV}_{11}=\frac{L^2}{8\pi\left(L^2+r^2\right)^{3/2} \left(2 m-\sqrt{L^2+r^2}\right)},\label{SETSV1}\\
&&T^{SV}_{22}=\frac{L^2 \left(\sqrt{L^2+r^2}-m\right)}{8\pi\left(L^2+r^2\right)^{3/2}},\qquad T^{SV}_{33}=\frac{L^2 \sin ^2\theta  \left(\sqrt{L^2+r^2}-m\right)}{8\pi\left(L^2+r^2\right)^{3/2}},\label{SETSV2}
\end{eqnarray}
and $T^{NMC}_{\mu\nu}$ is the stress-energy tensor that provides information about the nonminimal coupling between the Simpson--Visser solution and the cloud of string, which is given by
\begin{eqnarray}
   && T^{NMC}_{00}=-\frac{a L^2 \left((a-2) \sqrt{L^2+r^2}+6 m\right)}{8\pi\left(L^2+r^2\right)^{5/2}},\label{SETCP1}\\
    &&T^{NMC}_{11}=\frac{2 a L^2 m}{8\pi\left(L^2+r^2\right)^{3/2} \left(\sqrt{L^2+r^2}-2 m\right) \left((a-1) \sqrt{L^2+r^2}+2 m\right)},\label{SETCP2}\\
    &&T^{NMC}_{22}=-\frac{a L^2}{8\pi\left(L^2+r^2\right)},\qquad T^{CP}_{33}=-\frac{a L^2 \sin ^2\theta }{8\pi\left(L^2+r^2\right)}\label{SETCP3}.
\end{eqnarray}
If $a=0$ or $L=0$ all components of $T^{NMC}_{\mu\nu}$ are zero.

To the Simpson--Visser solution we have $\mathcal{R}=\sqrt{L^2+r^2}$ \cite{Simpson:2018tsi}. As the presence of a cloud of strings does not modify the area, we will consider the same $\mathcal{R}$. With this modification in the area, the equations related to the string are
\begin{equation}
\Sigma^{01}=\sqrt{-\gamma}=\frac{a}{\rho \left(r^2+L^2\right)}.    
\end{equation}
This result implies a modification in the stress-energy tensor to a cloud of string, whose nonzero components are given by
\begin{equation}
    {{T^{CS}}^{0}}_0={{T^{CS}}^{1}}_1=\frac{a}{8\pi \left(r^2+L^2\right)}.\label{SETstringmodified}
\end{equation}
Again, to guarantee the positivity of the energy density, we impose $a>0$.

Using the stress-energy tensors \eqref{SETSV1}, \eqref{SETSV2}, \eqref{SETCP1}--\eqref{SETCP3}, and \eqref{SETstringmodified}, the field equations are given by
\begin{eqnarray}
   &&-\frac{a L^2 \left((a-2) L^2+(a-2) r^2+6 m \sqrt{L^2+r^2}\right)}{(a-1) \left(L^2+r^2\right)^2+2 m \left(L^2+r^2\right)^{3/2}}-a-\frac{\left(L^2+r^2\right) \left(r
   f'-1\right)+f\left(2 L^2+r^2\right)}{\left(L^2+r^2\right)}\nonumber\\
   &&+\frac{L^2 \left(\sqrt{L^2+r^2}-4 m\right) \left(\sqrt{L^2+r^2}-2 m\right)}{f \left(L^2+r^2\right)^2}=0,\\
  && L^2 \left(-\frac{2 a m}{\sqrt{L^2+r^2}-2 m}-a+1\right)+(1-a) r^2-r \left(L^2+r^2\right) f'+ \frac{fL^2\left(L^2+r^2\right)^{1/2}}{
   \left(2 m-\sqrt{L^2+r^2}\right)}-f r^2=0,\\
  && -\frac{a L^2}{\left(L^2+r^2\right)^2}-\frac{r f'(r)}{L^2+r^2}-\frac{f''(r)}{2}-\frac{L^2 f(r)}{\left(L^2+r^2\right)^2}-\frac{L^2
   m}{\left(L^2+r^2\right)^{5/2}}+\frac{L^2\left(1-a\right)}{\left(L^2+r^2\right)^2}=0.
\end{eqnarray}
Integrating these equations, we find
\begin{equation}
    f(r)=1-a-\frac{2m}{\sqrt{L^2+r^2}}.\label{fSV}
\end{equation}
It is the metric coefficient that describes the Simpson--Visser solution immersed in a cloud of strings. If $L=0$, the Letelier solution is recovered, and if $a=0$, the Simpson--Visser is recovered. If $a=1$, there is no event horizon, so we impose $0<a<1$. The Kretschmann scalar associated with this solution is
\begin{eqnarray}
  && K(r)=\frac{4 \left(a^2 r^6+L^4 \left((2 (a-1) a+3) r^2+9 m^2\right)+L^2 \left(a (a+2) r^4-12 m^2 r^2\right)+(2 (a-2) a+3) L^6+12 m^2 r^4\right)}{\left(L^2+r^2\right)^5}\nonumber\\
   &&\qquad+\frac{16 m
   \left(-(a-2) L^2 r^2+2 (a-1) L^4+a r^4\right)}{\left(L^2+r^2\right)^{9/2}}
\end{eqnarray}
Expanding the Kretschmann scalar to $r\rightarrow 0$ and $r\rightarrow \infty$, we find
\begin{eqnarray}
K &\approx& \frac{8}{L^4} \left(a+\frac{2 m}{\sqrt{L^2}}-1\right)^2+\frac{4 m^2}{L^6}+\frac{4}{L^4},\ \mbox{to} \ r\rightarrow 0,\\
K &\approx& \frac{4 a^2}{r^4}+\frac{16 a m}{r^5}-\frac{8 \left(2 a^2 L^2-a L^2-6 m^2\right)}{r^6},\ \mbox{to} \ r\rightarrow \infty.
\end{eqnarray}
Different from the case before, this solution is always regular. What allows for regularity is the modification in the area.

The energy conditions, where $t$ is the timelike coordinate, are written as

\begin{eqnarray}
&&NEC_{1} \Longleftrightarrow
 \frac{2 L^2 \left((a-1) \sqrt{L^2+r^2}+2 m\right)}{\kappa ^2 \left(L^2+r^2\right)^{5/2}}\geq0, \qquad \qquad NEC_2 \Longleftrightarrow
\frac{a}{\kappa ^2\left(L^2+r^2\right)}+\frac{3 L^2 m}{\kappa ^2 \left(L^2+r^2\right)^{5/2}}\geq 0,\label{Sol2cond1+}\\
&&WEC_3 \Longleftrightarrow
\frac{(2 a-1) L^2+a r^2}{\kappa ^2 \left(L^2+r^2\right)^2}+\frac{4 L^2 m}{\kappa ^2 \left(L^2+r^2\right)^{5/2}}\geq0, \qquad \quad SEC_3 \Longleftrightarrow\frac{2 L^2 m}{\kappa ^2 \left(L^2+r^2\right)^{5/2}}\geq0,\label{Sol2cond2+}\\
&&DEC_{1} \Longrightarrow
 \frac{2 a}{\kappa ^2 \left(L^2+r^2\right)}+\frac{4 L^2 m}{\kappa ^2 \left(L^2+r^2\right)^{5/2}}\geq 0,\quad  DEC_2 \Longrightarrow
\frac{(2-3 a) L^2-a r^2}{\kappa ^2 \left(L^2+r^2\right)^2}-\frac{5 L^2 m}{\kappa ^2 \left(L^2+r^2\right)^{5/2}}\geq0. \label{Sol2cond3+}
\end{eqnarray}

Where $t$ is the spacelike coordinate, the energy conditions are
\begin{eqnarray}
&&NEC_{1} \Longleftrightarrow -\frac{2 (a-1) L^2}{\kappa ^2 \left(L^2+r^2\right)^2}-\frac{4 L^2 m}{\kappa ^2 \left(L^2+r^2\right)^{5/2}}\geq
 0,\qquad\qquad NEC_2\Longleftrightarrow
-\frac{L^2 m \left((2-a) L^2+a r^2\right)}{\kappa ^4 \left(L^2+r^2\right)^{9/2}}\geq 0,
\label{Sol2cond1-}\\
&&WEC_3 \Longleftrightarrow
\frac{a r^2+L^2}{\kappa ^2 \left(L^2+r^2\right)^2} \geq 0,\qquad \qquad SEC_3 \Longleftrightarrow
\frac{4 (a-1) L^2}{\kappa ^2 \left(L^2+r^2\right)^2}+\frac{6 L^2 m}{\kappa ^2 \left(L^2+r^2\right)^{5/2}} \geq0,
\label{Sol2cond2-}\\
&&DEC_{1} \Longrightarrow
\frac{2 a}{\kappa ^2 \left(L^2+r^2\right)}+\frac{4 L^2 m}{\kappa ^2 \left(L^2+r^2\right)^{5/2}}\geq 0, \qquad DEC_2 \Longrightarrow
\frac{a}{\kappa ^2 \left(L^2+r^2\right)}+\frac{L^2 m}{\kappa ^2 \left(L^2+r^2\right)^{5/2}}\geq 0.
\label{Sol2cond3-}
\end{eqnarray}
The null energy condition is violated outside the event horizon, which implies that the other energy conditions are also violated. It is important to emphasize that the energy density, to the Simpson--Visser spacetime, is negative in some region outside the event horizon. However, to $1>a\geq 1/2$ the energy density, $\rho = DEC_3=WEC_3$, is always positive.

\begin{figure}
    \centering
    \includegraphics[scale=0.9]{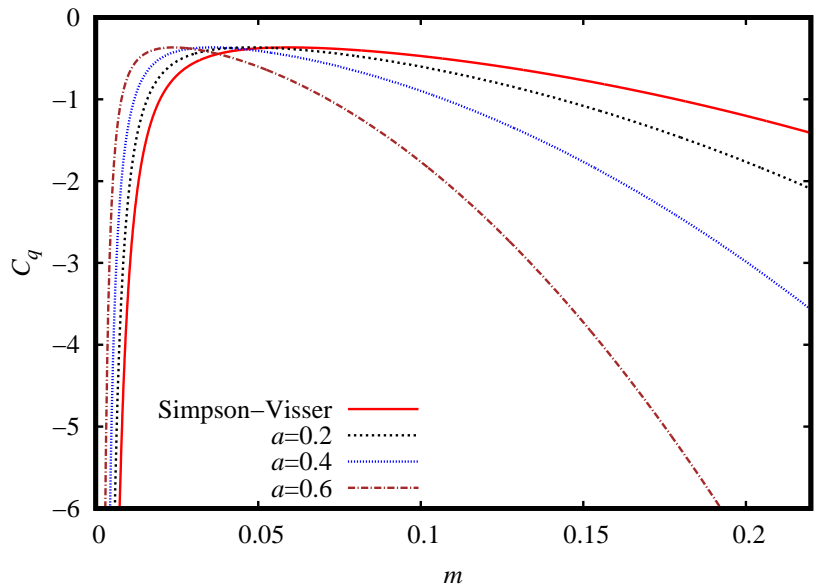}
    \caption{Heat capacity at constant charge to the solution Eq. \eqref{fSV} with $L=0.1$.}
    \label{fig:Cq2}
\end{figure}

From Eq. \eqref{Temp1}, the temperature is
\begin{equation}
T=\frac{m r}{2 \pi  \left(L^2+r_+^2\right)^{3/2}}. 
\end{equation}
Using the condition $f(r_+)=0$, the black hole mass is given by
\begin{equation}
    m=\frac{1}{2} (1-a) \sqrt{L^2+r_+^2},\label{massSV}
\end{equation}
and the temperature becomes
\begin{equation}
    T=\frac{r_+\left(1-a\right)}{4 \pi  L^2+4 \pi  r_+^2}.\label{TempSV}
\end{equation}
The entropy to this case is
\begin{equation}
    S=\pi \left(r_+^2+L^2\right).
\end{equation}
Once we have the temperature, entropy, and $r_+=\frac{\sqrt{(1-a)^2 L^2+4 m^2}}{1-a}$, the heat capacity at constant charge is
\begin{equation}
    C_q=-\frac{\pi  \left((a-1)^2 L^2+2 m^2\right) \left((a-1)^2 L^2+4 m^2\right)}{(a-1)^2 m^2}.\label{HCSV}
\end{equation}
 In Fig. \ref{fig:Cq2} we see that the heat capacity is always negative, $C_q<0$, and there is no phase transition.

\section{Conclusion}\label{S:conclusion}

In this work, we analyze the consequences of inserting regular black holes into a cloud of strings in the context of general relativity, studying the regularity, energy conditions, and thermodynamics of these solutions. For this, we first found the stress-energy tensor of a cloud of strings.

We consider the Einstein equations together with the stress-energy tensor associated with the Bardeen solution and the stress-energy tensor of the string cloud. By integrating the equations of motion, we were able to obtain the Bardeen solution surrounded by a kind of string atmosphere. This solution presents a problem when considering regularity. The presence of a cloud makes the solution singular, so the geodesics are interrupted at $r=0$. The presence of the atmosphere changes the energy conditions but not significantly. For instance, if the parameter $a$ does not appear in $SEC_3$, the strong energy condition is violated inside the event horizon, as it happens in the standard Bardeen solution. The null energy condition is satisfied and the energy density, $WEC_3$, is positive. We noticed that the presence of the cloud affects the heat capacity, decreasing the region where the thermodynamically stable phase exists, $C_q>0$. Despite the modifications in thermodynamics, the solution presents two phase transitions, in the same way as for Bardeen, from unstable to stable and then from stable to unstable.

To obtain the Simpson--Visser solution surrounded by a cloud of strings, it was necessary to consider the stress-energy tensor for a cloud of strings, for the Simpson--Visser spacetime, and also to introduce a new stress-energy tensor that describes the interaction between the cloud and the black hole. The interaction term disappears if $a=0$ or $L=0$. The solution was obtained through the integration of the Einstein equations. To $L=0$ the Letelier solution is recovered, and to $a=0$ the usual Simpson--Viser solution is recovered. Different from the usual regular black hole, solutions that do not have their area modified, the Simpson--Visser solution is regular even when immersed in the cloud. The parameter $a$ does not appear only in $SEC_3$, which means that the gravitational interaction is always attractive. To $a\geq1/2$ the energy density is always positive but if $a\geq 1$ there is no horizon. So, if $a=1/2$, the event horizon radius is $r_H=\sqrt{16m^2-L^2}$. Regardless of the value of $L$, the energy density will be positive, so we can have a black bounce or a wormhole with positive energy density. The thermodynamics of this solution is not drastically affected by the presence of the cloud, since, regardless of the value of $a$, the solution is always thermodynamically unstable, $C_q<0$.

In future work, we hope to investigate more properties of these spacetimes, such as the behavior of particles/fields around them.


\section*{Acknowledgements}
M.E.R.  thanks Conselho Nacional de Desenvolvimento Cient\'ifico e Tecnol\'ogico - CNPq, Brazil  for partial financial support. This study was financed in part by the Coordena\c{c}\~ao de Aperfei\c{c}oamento de Pessoal de N\'ivel Superior - Brasil (CAPES) - Finance Code 001.




\begin{thebibliography}{99}

\bibitem{din} R. D'Inverno, ``Introducing Einstein's Relativity'', Oxford University Press, New York (1998).

\bibitem{wal} R. M. Wald, ``General Relativity'', The University of Chicago Press, Chicago (1984).


\bibitem{Einstein:1915bz}
A.~Einstein,
``Explanation of the Perihelion Motion of Mercury from the General Theory of Relativity'',
Sitzungsber. Preuss. Akad. Wiss. Berlin (Math. Phys. ) \textbf{1915}, 831-839 (1915).



\bibitem{Kraniotis:2003ig}
G.~V.~Kraniotis and S.~B.~Whitehouse,
``Exact calculation of the perihelion precession of mercury in general relativity, the cosmological constant and jacobi's inversion problem'',
Class. Quant. Grav. \textbf{20}, 4817-4835 (2003),
[arXiv:astro-ph/0305181 [astro-ph]].


\bibitem{Will:2018mcj}
C.~M.~Will,
``New General Relativistic Contribution to Mercury\textquoteright{}s Perihelion Advance'',
Phys. Rev. Lett. \textbf{120}, no.19, 191101 (2018),
[arXiv:1802.05304 [gr-qc]].

\bibitem{Crispino:2019yew}
L.~C.~B.~Crispino and D.~Kennefick,
``100 years of the first experimental test of General Relativity'',
Nature Phys. \textbf{15}, 416 (2019),
[arXiv:1907.10687 [physics.hist-ph]].

\bibitem{LIGOScientific:2016aoc}
B.~P.~Abbott \textit{et al.} [LIGO Scientific and Virgo],
``Observation of Gravitational Waves from a Binary Black Hole Merger'',
Phys. Rev. Lett. \textbf{116}, no.6, 061102 (2016),
[arXiv:1602.03837 [gr-qc]].

\bibitem{Chandrasekhar:1985kt}
S.~Chandrasekhar,
``The mathematical theory of black holes'', Published in: Oxford University Press, New York (1992).




\bibitem{LIGOScientific:2018mvr}
B.~P.~Abbott \textit{et al.} [LIGO Scientific and Virgo],
``GWTC-1: A Gravitational-Wave Transient Catalog of Compact Binary Mergers Observed by LIGO and Virgo during the First and Second Observing Runs'',
Phys. Rev. X \textbf{9}, no.3, 031040 (2019),
[arXiv:1811.12907 [astro-ph.HE]].


\bibitem{LIGOScientific:2020ibl}
R.~Abbott \textit{et al.} [LIGO Scientific and Virgo],
``GWTC-2: Compact Binary Coalescences Observed by LIGO and Virgo During the First Half of the Third Observing Run'',
Phys. Rev. X \textbf{11}, 021053 (2021),
[arXiv:2010.14527 [gr-qc]].

\bibitem{LIGOScientific:2021psn}
R.~Abbott \textit{et al.} [LIGO Scientific, VIRGO and KAGRA],
``The population of merging compact binaries inferred using gravitational waves through GWTC-3'',
[arXiv:2111.03634 [astro-ph.HE]].



\bibitem{Stoica:2014tpa}
O.~C.~Stoica,
``The Geometry of Black Hole singularities'',
Adv. High Energy Phys. \textbf{2014}, 907518 (2014),
[arXiv:1401.6283 [gr-qc]].


\bibitem{Bronnikov:2012wsj}
K.~A.~Bronnikov and S.~G.~Rubin,
``Black Holes, Cosmology and Extra Dimensions'', World
Scientific, Singapore (2013).



\bibitem{Ansoldi:2008jw}
S.~Ansoldi,
``Spherical black holes with regular center: A Review of existing models including a recent realization with Gaussian sources'',
[arXiv:0802.0330 [gr-qc]].



\bibitem{Bardeen} 
J. M. Bardeen,{\it{ Non-singular general relativistic gravitational collapse}}, in Proceedings of the International Conference GR5, Tbilisi, U.S.S.R. (1968).

\bibitem{Beato1}
E.~Ayon-Beato and A.~Garcia,
``The Bardeen model as a nonlinear magnetic monopole'',
Phys. Lett. B \textbf{493}, 149-152 (2000),
[arXiv:gr-qc/0009077 [gr-qc]].


\bibitem{Fan-Wang}
Z.~Y.~Fan and X.~Wang, ``Construction of Regular Black Holes in General Relativity'', \\
Phys. Rev. D {\bf 94}, no.12, 124027 (2016), 
[\href{https://arxiv.org/abs/1610.02636}{arXiv:1610.02636} [gr-qc]].

\bibitem{Zaslavskii}
O.~B.~Zaslavskii, ``Regular black holes and energy conditions'', Phys. Lett. B {\bf 688}, 278-280 (2010),
[\href{https://arxiv.org/abs/1004.2362}{arXiv:1004.2362} [gr-qc]].

\bibitem{Rodrigues:2015}
M.~E.~Rodrigues, E.~L.~B.~Junior, G.~T.~Marques and V.~T.~Zanchin,\\
``Regular black holes in $f(R)$ gravity coupled to nonlinear electrodynamics'',\\
Phys. Rev. D \textbf{94} (2016) no.2, 024062,
[arXiv:1511.00569 [gr-qc]].


\bibitem{Rodrigues:2017}
M.~E.~Rodrigues, E.~L.~B.~Junior and M.~V.~de S.~Silva,\\
``Using dominant and weak energy conditions for building new classes of regular black holes'',\\
JCAP \textbf{02} (2018), 059,
[arXiv:1705.05744 [physics.gen-ph]].

\bibitem{Rodrigues:2018}
M.~E.~Rodrigues and M.~V.~d.~Silva,
``Bardeen Regular Black Hole With an Electric Source'',
JCAP \textbf{06} (2018), 025,
[arXiv:1802.05095 [gr-qc]].

\bibitem{Rodrigues:2019}
M.~E.~Rodrigues and M.~V.~de S.~Silva,
``Regular multi-horizon black holes in $f(G)$ gravity with nonlinear electrodynamics'',
Phys. Rev. D \textbf{99} (2019) no.12, 124010,
[arXiv:1906.06168 [gr-qc]].

\bibitem{Silva:2018}
M.~V.~d.~Silva and M.~E.~Rodrigues,
``Regular black holes in $f(G)$ gravity'',
Eur. Phys. J. C \textbf{78} (2018) no.8, 638,
[arXiv:1808.05861 [gr-qc]].

\bibitem{Junior:2020}
E.~L.~B.~Junior, M.~E.~Rodrigues and M.~V.~d.~S.~Silva,
``Regular Black Holes in Rainbow Gravity'',\\{}
[arXiv:2002.04410 [gr-qc]].



\bibitem{Bronnikov:2017tnz}
K.~A.~Bronnikov,
``Comment on \textquotedblleft{}Construction of regular black holes in general relativity\textquotedblright{}'',
Phys. Rev. D \textbf{96}, no.12, 128501 (2017),
[\href{https://arxiv.org/abs/1712.04342}{arXiv:1712.04342} [gr-qc]].



\bibitem{NED2}
E. Ay\'on-Beato and A. Garc\'{\i}a, ``New regular black hole solution 
from nonlinear electrodynamics'', Phys.Lett. B {\bf 464}, 25 (1999), 
[\href{http://arxiv.org/abs/hep-th/9911174}{arXiv:9911174} [hep-th]].

\bibitem{NED3}
E. Ay\'on-Beato and A. Garc\'\i a, ``Regular black hole in general 
relativity coupled to nonlinear electrodynamics'', Phys. Rev. Lett. {\bf 80}, 
5056 (1998), [\href{http://arxiv.org/abs/gr-qc/9911046}{arXiv:9911046} [gr-qc]].

\bibitem{NED4}
K. A. Bronnikov, ``Regular magnetic black holes 
and monopoles from nonlinear electrodynamics'', Phys. Rev. D {\bf 63}, 044005 
(2001), [\href{http://arxiv.org/abs/gr-qc/0006014}{arXiv:0006014} [gr-qc]].

\bibitem{NED5}
I. Dymnikova, ``Regular electrically charged structures in nonlinear 
electrodynamics coupled to general relativity'', Classical Quantum Gravity 
{\bf 21}, 4417 (2004), [\href{http://arxiv.org/abs/gr-qc/0407072}{arXiv:0407072} [gr-qc]].


\bibitem{Bronnikov:2006fu}
K.~A.~Bronnikov, V.~N.~Melnikov and H.~Dehnen,
``Regular black holes and black universes'',
Gen. Rel. Grav. \textbf{39}, 973-987 (2007),
[\href{https://arxiv.org/abs/gr-qc/0611022}{arXiv:gr-qc/0611022} [gr-qc]].



\bibitem{NED10}
L. Balart and E. C. Vagenas, ``Regular black holes with a nonlinear 
electrodynamics source'', Phys. Rev. D {\bf 90}, 124045 (2014),
[\href{http://arxiv.org/abs/arXiv:1408.0306}{arXiv:1408.0306} [gr-qc]].



\bibitem{neves} 
J.~C.~S.~Neves, A. Saa, ``Regular rotating black holes and the weak energy condition'', 
Phys. Lett. B \textbf{734} (2014), 44-48, 
[\href{https://arxiv.org/abs/1402.2694}{arXiv:1402.2694} [gr-qc]].

\bibitem{toshmatov} 
B. Toshmatov, B. Ahmedov, A. Abdujabbarov, Z. Stuchlik, ``Rotating Regular Black Hole Solution'', 
Phys. Rev. D \textbf{89} (2014) no. 10, 104017, 
[\href{https://arxiv.org/abs/1404.6443}{arXiv:1404.6443} [gr-qc]].



\bibitem{ramon} 
R. Torres, F. Fayos, ``On regular rotating black holes'', 
Gen. Rel. Grav. \textbf{49}, no.1, 2 (2017)
[\href{https://arxiv.org/abs/1611.03654}{arXiv:1611.03654} [gr-qc]].

\bibitem{berej} 
W. Berej, J. Matyjasek, D. Tryniecki, M. Woronowicz, ``Regular black holes in quadratic gravity'', 
Gen. Rel. Grav. \textbf{38} (2006) 885-906 ,
[\href{https://arxiv.org/abs/hep-th/0606185}{arXiv:0606185} [hep-th]].


\bibitem{Simpson:2018tsi}
A.~Simpson and M.~Visser,
``Black-bounce to traversable wormhole'',
JCAP \textbf{02} (2019), 042,
[\href{https://arxiv.org/abs/1812.07114}{arXiv:1812.07114} [gr-qc]].

\bibitem{Bronnikov:2021uta}
K.~A.~Bronnikov and R.~K.~Walia,
``Field sources for Simpson-Visser spacetimes'',
Phys. Rev. D \textbf{105}, no.4, 044039 (2022),
[arXiv:2112.13198 [gr-qc]].


\bibitem{Canate:2022gpy}
P.~Ca\~nate,
``Black-bounces as magnetically charged phantom regular black holes in Einstein-nonlinear electrodynamics gravity coupled to a self-interacting scalar field'',
[arXiv:2202.02303 [gr-qc]].


\bibitem{Simpson:2019cer}
A.~Simpson, P.~Mart\'in-Moruno and M.~Visser,
``Vaidya spacetimes, black-bounces, and traversable wormholes'',\\
Class. Quant. Grav. \textbf{36} (2019) no.14, 145007,
[\href{https://arxiv.org/abs/1902.04232}{arXiv:1902.04232} [gr-qc]].




\bibitem{Lobo:2020ffi}
F.~S.~N.~Lobo, M.~E.~Rodrigues, M.~V.~d.~S.~Silva, A.~Simpson and M.~Visser,
``Novel black-bounce spacetimes: wormholes, regularity, energy conditions, and causal structure'',
Phys. Rev. D \textbf{103}, no.8, 084052 (2021),
[arXiv:2009.12057 [gr-qc]].








\bibitem{Simpson:2021vxo}
A.~Simpson,
``From black-bounce to traversable wormhole, and beyond'',
[arXiv:2110.05657 [gr-qc]].







\bibitem{Lobo:2020kxn}
F.~S.~N.~Lobo, A.~Simpson and M.~Visser,
``Dynamic thin-shell black-bounce traversable wormholes'',\\
Phys. Rev. D \textbf{101} (2020) no.12, 124035
doi:10.1103/PhysRevD.101.124035
[\href{https://arxiv.org/abs/2003.09419}{arXiv:2003.09419} [gr-qc]].








\bibitem{Tsukamoto:2021caq}
N.~Tsukamoto,
``Gravitational lensing by two photon spheres in a black-bounce spacetime in strong deflection limits'',
Phys. Rev. D \textbf{104}, no.6, 064022 (2021),
[arXiv:2105.14336 [gr-qc]].

\bibitem{Olmo:2021piq}
G.~J.~Olmo, D.~Rubiera-Garcia and D.~S.~C.~G\'omez,
``New light rings from multiple critical curves as observational signatures of black hole mimickers'',
[arXiv:2110.10002 [gr-qc]].

\bibitem{Guerrero:2022qkh}
M.~Guerrero, G.~J.~Olmo, D.~Rubiera-Garcia and D.~S.~C.~G\'omez,
``Light ring images of double photon spheres in black hole and wormhole space-times'',
[arXiv:2202.03809 [gr-qc]].

\bibitem{Yang:2021cvh}
Y.~Yang, D.~Liu, Z.~Xu, Y.~Xing, S.~Wu and Z.~W.~Long,
``Echoes of novel black-bounce spacetimes,''
Phys. Rev. D \textbf{104}, no.10, 104021 (2021),
[arXiv:2107.06554 [gr-qc]].

\bibitem{Lima:2020auu}
H.~C.~D.~Lima, C.~L.~Benone and L.~C.~B.~Crispino,
``Scalar absorption: Black holes versus wormholes'',
Phys. Rev. D \textbf{101}, no.12, 124009 (2020),
doi:10.1103/PhysRevD.101.124009
[arXiv:2006.03967 [gr-qc]].

\bibitem{Lima:2021las}
H.~C.~D.~Lima, Junior., L.~C.~B.~Crispino, P.~V.~P.~Cunha and C.~A.~R.~Herdeiro,
``Can different black holes cast the same shadow?'',
Phys. Rev. D \textbf{103}, no.8, 084040 (2021),
[arXiv:2102.07034 [gr-qc]].

\bibitem{Huang:2019arj}
H.~Huang and J.~Yang,
``Charged Ellis Wormhole and Black Bounce'',
Phys. Rev. D \textbf{100}, no.12, 124063 (2019),
[arXiv:1909.04603 [gr-qc]].



\bibitem{Rodrigues:2022mdm}
M.~E.~Rodrigues and M.~V.~d.~S.~Silva,
``Black Bounces with multiple throats and anti-throats'',
[arXiv:2204.11851 [gr-qc]].


\bibitem{Letelier:1979ej}
P.~S.~Letelier,
``CLOUDS OF STRINGS IN GENERAL RELATIVITY'',
Phys. Rev. D \textbf{20}, 1294-1302 (1979).

\bibitem{Glass:1997qz}
E.~N.~Glass and J.~P.~Krisch,
``Radiation and string atmosphere for relativistic stars'',
Phys. Rev. D \textbf{57}, 5945-5947 (1998),
[arXiv:gr-qc/9803040 [gr-qc]].


\bibitem{Ganguly:2014cqa}
A.~Ganguly, S.~G.~Ghosh and S.~D.~Maharaj,
``Accretion onto a black hole in a string cloud background'',
Phys. Rev. D \textbf{90}, no.6, 064037 (2014),
[arXiv:1409.7872 [gr-qc]].


\bibitem{deMToledo:2018tjq}
J.~de M.Toledo and V.~B.~Bezerra,
``Black holes with cloud of strings and quintessence in Lovelock gravity'',
Eur. Phys. J. C \textbf{78}, no.7, 534 (2018).

\bibitem{Toledo:2018hav}
J.~M.~Toledo and V.~B.~Bezerra,
``The Reissner\textendash{}Nordstr\"om black hole surrounded by quintessence and a cloud of strings: Thermodynamics and quasinormal modes'',
Int. J. Mod. Phys. D \textbf{28}, no.01, 1950023 (2018).


\bibitem{MoraisGraca:2016hmv}
J.~P.~Morais Gra\c{c}a, G.~I.~Salako and V.~B.~Bezerra,
``Quasinormal modes of a black hole with a cloud of strings in Einstein\textendash{}Gauss\textendash{}Bonnet gravity'',
Int. J. Mod. Phys. D \textbf{26}, no.10, 1750113 (2017),
[arXiv:1604.04734 [gr-qc]].

\bibitem{DiaseCosta:2018xyj}
M.~M.~Dias e Costa, J.~M.~Toledo and V.~B.~Bezerra,
``The Letelier spacetime with quintessence: Solution, thermodynamics and Hawking radiation'',
Int. J. Mod. Phys. D \textbf{28}, no.06, 1950074 (2019),
[arXiv:1811.12585 [gr-qc]].

\bibitem{Chabab:2020ejk}
M.~Chabab and S.~Iraoui,
``Thermodynamic criticality of d-dimensional charged AdS black holes surrounded by quintessence with a cloud of strings background'',
Gen. Rel. Grav. \textbf{52}, no.8, 75 (2020),
[arXiv:2001.06063 [hep-th]].


\bibitem{Cai:2019nlo}
X.~C.~Cai and Y.~G.~Miao,
``Quasinormal modes and spectroscopy of a Schwarzschild black hole surrounded by a cloud of strings in Rastall gravity'',
Phys. Rev. D \textbf{101}, no.10, 104023 (2020),
[arXiv:1911.09832 [hep-th]].

\bibitem{Ghosh:2014pga}
S.~G.~Ghosh, U.~Papnoi and S.~D.~Maharaj,
``Cloud of strings in third order Lovelock gravity'',
Phys. Rev. D \textbf{90}, no.4, 044068 (2014),
[arXiv:1408.4611 [gr-qc]].

\bibitem{Ghosh:2014dqa}
S.~G.~Ghosh and S.~D.~Maharaj,
``Cloud of strings for radiating black holes in Lovelock gravity'',
Phys. Rev. D \textbf{89}, no.8, 084027 (2014),
[arXiv:1409.7874 [gr-qc]].



\bibitem{Sood:2022fio}
A.~Sood, A.~Kumar, J.~K.~Singh and S.~G.~Ghosh,
``Thermodynamic stability and $P-V$ criticality of nonsingular-AdS black holes endowed with clouds of strings'',
Eur. Phys. J. C \textbf{82}, no.3, 227 (2022),
[arXiv:2204.05996 [gr-qc]].

\bibitem{book}
M. Visser, \textit{Lorentzian wormholes: From Einstein to Hawking}, AIP press [now Springer], New York (1995).

\bibitem{Azreg-Ainou:2012zyx}
M.~Azreg-A\"\i{}nou and M.~E.~Rodrigues,
``Thermodynamical, geometrical and Poincar\'e methods for charged black holes in presence of quintessence'',
JHEP \textbf{09}, 146 (2013),
[arXiv:1211.5909 [gr-qc]].

\end{thebibliography}
\end{document}